\documentclass[twoside]{dis07}
\usepackage[latin1]{inputenc}
\usepackage[dvips]{graphicx,epsfig,color}
\usepackage{wrapfig,rotating}
\usepackage{amssymb,amsmath,array}

\begin{document}
\title{New QCD fits to HERA data and search for exclusive events at the Tevatron}

%***********************************************************************
% AUTHORS INFORMATION AREA
%***********************************************************************
\author{Old\v{r}ich Kepka and Christophe Royon
%
% Optional short acknowledgment: remove next line if non-needed
%\thanks{On behalf of the RP220 Collaboration}
%
% DO NOT MODIFY THE FOLLOWING '\vspace' ARGUMENT
\vspace{.3cm}\\
%
% Addresses and institutions (remove "1- " in case of a single institution)
DAPNIA/Service de physique des particules, \\ CEA/Saclay, 91191 
Gif-sur-Yvette cedex, France
%
% Remove the next three lines in case of a single institution
}
%***********************************************************************
% END OF AUTHORS INFORMATION AREA
%***********************************************************************

\maketitle

\begin{abstract}
We describe new QCD fits to diffractive proton structure functions measured at HERA,
and we use these parton densities to predict the shape of the dijet mass
fraction at the Tevatron and look for the existence of exclusive events in the
dijet channel.
\end{abstract}

\section{QCD fits to proton diffractive structure function data from HERA}
\begin{figure}
\centerline{\includegraphics[width=0.7\columnwidth]{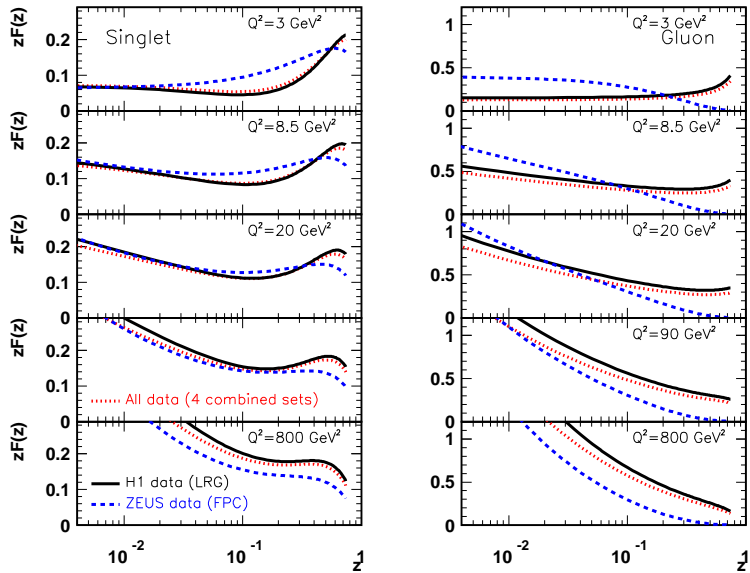}}
\caption{Gluon and quark densities in the pomeron measured using H1 and ZEUS
data.}\label{Fig1}
\end{figure}
We use the most recent published data \cite{h1zeus} on diffractive proton
structure function measured by the H1 and ZEUS collaborations. Data are fitted
using the following quark and gluon densities \cite{pdf}: 
\begin{eqnarray}
z{\it {S}}(z,Q^2=Q_0^2) &=& \left[
A_S z^{B_S}(1-z)^{C_S} (1+D_S z + E_S \sqrt{z} ) \right] %\nonumber \\
\cdot e^{\frac{0.01}{z-1}} \label{quarka} \nonumber \\
z{\it {G}}(z,Q^2=Q_0^2) &=& \left[
A_G (1-z)^{C_G}   \right]
\cdot e^{\frac{0.01}{z-1}}. \nonumber
\end{eqnarray}
In the fits, $\alpha_S(M_Z)=0.18$ and the initial scale is taken at $Q_0^2=3$
GeV$^2$. The charm quark contribution is computed in the fixed flavour scheme
using the photon-gluon fusion prescription. The pomeron intecept is found to be
0.12 using H1 data and $\chi^2/dof \sim 0.9$. With respect to the ``standard" H1
approach for the QCD fits, we have more parameters for the quark and gluon
densities at the starting scale which allows to fix the starting scale at 3
GeV$^2$ and not to fit it. We cross checked that we find the same results as H1
while making the same assumptions. Other approaches based on dipole and
saturation models \cite{dipole} were also tested in Ref. \cite{pdf}.

%\begin{wrapfigure}{r}{0.5\columnwidth}
%\end{wrapfigure}

The gluon and quark densities are given in
Fig. 1. While the quark densities are found to be relatively close for H1 and
ZEUS, the gluon density differs by more than a factor 2. New preliminary data from
ZEUS reduce this discrepancy. In the following, we will only use the QCD fits
to the H1 data to compare with the dijet mass fractions measured in the CDF
collaboration at the Tevatron.
It is also worth noticing that the gluon density is poorly known at high
$\beta$, where $\beta$ is the momentum fraction of the pomeron carried by the
interacting parton. To illustrate this, we multiply the gluon density by the
factor $(1-\beta)^{\nu}$ and fit the parameter $\nu$. The fit leads to $\nu =
0.0 \pm 0.6$ which demonstrates a large uncertainty of the gluon density at high
$\beta$ measured at HERA.

\section{Search for exclusive events at the Tevatron}

Exclusive events at the Tevatron or the LHC show the interesting property that
the full available energy in the pomeron-pomeron system for double pomeron
exchange events is used to produce the heavy mass object (dijet, diphoton...).
In other words, no energy is lost in pomeron remnants. Tagging both protons
scattered in the final state allow to measure precisely the kinematic
properties, for instance the mass, of the produced heavy object.
Exclusive events at the LHC recently captured high interest since it might be a
possibility to detect the Higgs boson diffractively by tagging the diffracted
protons in the final state \cite{higgs}.

\subsection{Search for exclusive events in $\chi_C$ production}
The CDF collaboration performed the search for exclusive events in the $\chi_C$
channel \cite{chic}. They obtained an upper limit of $\chi_C$ exclusive production
in the  $J/\Psi \gamma$ channel of $\sigma \sim 49$ pb $\pm$18 $\pm$ 39
pb for $y<0.6$.
In Ref. \cite{murilo}, we found that the contamination of
inclusive events into the signal region (the tail of the inclusive distribution
when little energy is taken away by the pomeron remnants) depends stronly on the
assumptions on the gluon distribution in the pomeron at high $\beta$ or in other
words on the $\nu$ parameter. Therefore, this channel is unfortunately not
conclusive concerning the existence of exclusive events.

\subsection{Search for exclusive events using the dijet mass fraction at the Tevatron}
One selects events with two jets only and one
looks at the dijet mass fraction distribution, the ratio between the dijet mass
and the total diffractive mass in the event. The CDF collaboration measured this
quantity for different jet $p_T$ cuts \cite{cdfrjj}. We compare this measurement
with different models of inclusive diffraction, namely ``factorised" (FM)
and ``Bialas Landshoff" (BL) models \cite{model}.
In the FM models, one takes the gluon and quark densities in the pomeron measured at HERA
as described in the previous section and the factorisation breaking between
HERA and the Tevatron only comes through the gap survival probability. The BL model
is non perturbative and diffraction is obtained via the exchange of a soft
pomeron, which means that the mass dependence of the exclusive cross section
is quite low.  The comparison between the CDF data for a jet $p_T$ cut of 10 GeV as an
example and the predictions from the FM model is given in Fig. 2. We also give
in the same figure the effects of changing the gluon density at high $\beta$ (by
changing the value of the $\nu$ parameter) and we note that inclusive
diffraction is not able to describe the CDF data at high dijet mass fraction,
where exclusive events are expected to appear \cite{olda}. The conclusion
remains unchanged when jets with $p_T>25$ GeV are considered \cite{olda}.

\begin{figure}
%\begin{wrapfigure}{r}{0.5\columnwidth}
\centerline{\includegraphics[width=0.47\columnwidth]{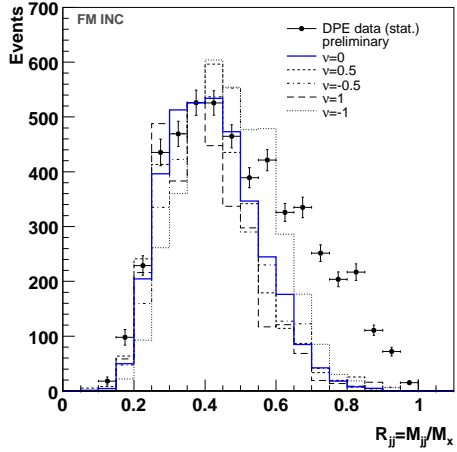}}
\caption{Dijet mass fraction measured by the CDF collaboration compared to the
prediction from the ``factorised model" for inclusive diffraction. The gluon
density in the pomeron at high $\beta$ was modified by varying the parameter
$\nu$.}
\label{Fig2}
%\end{wrapfigure}
\end{figure}

Adding exclusive events to the distribution of the dijet mass fraction leads to
a good description of data \cite{olda} as shown in Fig. 3 where we superimpose the
predictions from inclusive diffraction from the ``factorised" model and
exclusive one from the Durham model \cite{model}. It is worth noticing that the exclusive
``Bialas Landshoff" model \cite{model} leads to a too small dependence of the diffractive
exclusive cross section as a function of jet transverse momentum \cite{olda}.
In Ref. \cite{olda}, the CDF data were also compared to the soft colour
interaction models \cite{model}. While the need for exclusive events is less
obvious for this model, especially at high jet $p_T$, the jet rapidity
distribution measured by the CDF collaboration is badly reproduced. This is due
to the fact that, in the SCI model, there is a large difference between
requesting an intact proton in the final state and a rapidity gap \cite{olda}.

\subsection{Observation of exclusive events at the LHC}
The
exclusive contribution manifests itself as an increase in the tail of the dijet
mass fraction
distribution. Exclusive production slowly 
turns on with the increase of the jet $p_T$ (see Ref. \cite{olda}) and  
with respect to the uncertainty on the gluon density this appearance is 
almost negligible. The exclusive production at the LHC plays a minor role for 
low $p_T$ jets. Therefore, measurements e.g for $p_T<200\,\mathrm{GeV}$ 
where the inclusive production is dominant could be used 
to constrain the gluon density in the pomeron. The higher $p_T$ jet region can
be used to extract the exclusive 
contribution from the tail of the dijet mass fraction distribution. 
The extraction of the inclusive and exclusive jet production cross section will
be of great importance at the beginning of the LHC to be able to make precise
predictions on exclusive Higgs production and the background later on.

%\begin{wrapfigure}{r}{0.5\columnwidth}
%\end{wrapfigure}

%\begin{figure}
%\end{figure}
\begin{footnotesize}

% ----------------------------------------------------------------------------

%\begin{wrapfigure}{r}{0.5\columnwidth}
\begin{figure}
\centerline{\includegraphics[width=0.47\columnwidth]{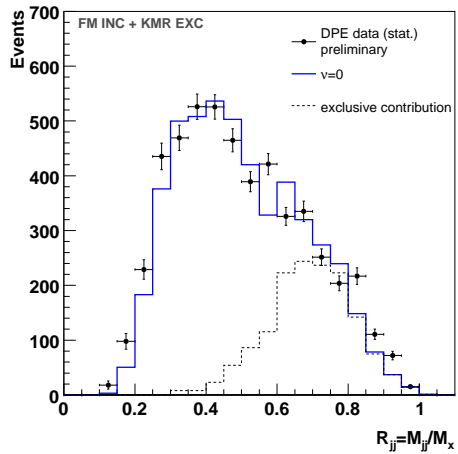}}
\caption{Dijet mass fraction measured by the CDF collaboration compared to the
prediction from ``factorised models" for inclusive diffraction and from the
Durham model for exclusive diffraction.}
\label{Fig3}
\end{figure}
%\end{wrapfigure}

\end{footnotesize}

% ****************************************************************************
% END OF BIBLIOGRAPHY AREA
% ****************************************************************************

\end{document}